\documentclass[aps,prb,twocolumn,preprintnumbers,amsmath,amssymb,superscriptaddress]{revtex4}%

\usepackage{graphicx}%
\usepackage{dcolumn}
\usepackage{amsmath}
\usepackage{color}

\begin{document}

\title{The single- {\it vs.} two-gap scenario: the specific heat and the thermodynamic critical field of BeAu superconductor }
%



\author{Rustem Khasanov}
 \email{rustem.khasanov@psi.ch}
 \affiliation{Laboratory for Muon Spin Spectroscopy, Paul Scherrer Institute, 5232 Villigen, Switzerland}

\author{Ritu Gupta}
 \affiliation{Laboratory for Muon Spin Spectroscopy, Paul Scherrer Institute, 5232 Villigen, Switzerland}

\author{Debarchan Das}
 \affiliation{Laboratory for Muon Spin Spectroscopy, Paul Scherrer Institute, 5232 Villigen, Switzerland}

\author{Andreas Leithe-Jasper}
 \affiliation{Max-Planck-Institut f\"{u}r Chemische Physik fester Stoffe, N\"{o}thnitzer Stra{\ss}e 40, 01187 Dresden, Germany}

\author{Eteri Svanidze}
 \affiliation{Max-Planck-Institut f\"{u}r Chemische Physik fester Stoffe, N\"{o}thnitzer Stra{\ss}e 40, 01187 Dresden, Germany}

\begin{abstract}
The puzzling situation where some thermodynamic quantities require a single-gap description, while others need a more complex gap scenario, is discussed. Our approach reveals that in some cases, the  conclusions based on measurements of only one thermodynamic quantity may lead to conflicting results. As an example, temperature evolutions of the electronic specific heat [$C_{\rm e}(T)$] and the thermodynamic critical field [$B_{\rm c}(T)$] of BeAu superconductor were reconsidered within the framework of the single-gap and the two-gap scenarios. The analysis shows that the single-gap approach describes the temperature dependencies of $C_{\rm e}(T)$ and the superfluid density $\rho_{\rm s}(T)$ satisfactorily. However it fails in the case of $B_{\rm c}(T)$. The self-consistent two-gap model, in contrast, is able to describe all of the mentioned thermodynamic quantities by using a similar set of parameters. Our results reveal that a proper description of the thermodynamic quantities, related to the superconducting pairing mechanism, requires the  use of similar model containing the same set of parameters.
\end{abstract}

\maketitle

Measurements of temperature dependencies of thermodynamic quantities such as, the superfluid density ($\rho_s$), the electronic specific heat ($C_e$), and the thermodynamic critical field ($B_{\rm c}$) are among the primary ways used to test pairing mechanisms of new superconductors.\cite{Shoenberg_book_1952, Parks_book_1969, Tinkham_book_1975, Tilley-Tilley_book_1990, de_Gennes_book_1999, Poole_book_2014} In particular, such experiments allow
 (i) to distinguish between the single-band and the multi-band scenario,\cite{Bouquet_EPL_2001, Carrington_2003, Guritanu_PRB_2004, Fletcher_PRL_2005, Prozorov_SST_2006, Khasanov_La214_PRL_2007, Khasanov_Y123_PRL_2007, Khasanov_Y123_JSNM_2008}
 (ii) to determine the superconducting gap structure,\cite{Khasanov_La214_PRL_2007, Khasanov_Y123_PRL_2007, Khasanov_Y123_JSNM_2008, Nakajima_PRL_2008, Evtushinsky_NJP_2009, Khasanov_Ba122_PRL_2009, Khasanov_Sr122_PRL_2009, Khasanov_Bi2201_PRL_2008, Gordon_PRB_2008, Khasanov_Bi2201_PRB_2009, Singh_PRB_2010, Khasanov_Bi2201_PRB_2010, Kim_PRB_2011, Chen_NJP_2013, Khasanov_SrPtP_PRB_2014, Khasanov_FeSeInt_PRB_2016, Verchenko_PRB_2017, Khasanov_1144_PRB_2018, Khasanov_1144_PRB_2019, Kim_Symmetry_2019}
 (iii) to calculate the temperature evolution(s) of the superconducting energy gaps,\cite{Singh_PRB_2010, Khasanov_Bi2201_PRB_2010, Kim_PRB_2011, Kim_Symmetry_2019, Khasanov_1144_PRB_2018, Khasanov_Bi-III_PRB_2018, Khasanov_1144_PRB_2019} {\it etc}.
Obviously, the parameters obtained from the analysis of various thermodynamic quantities need to be consistent with each other. In other words, the temperature evolution of the superfluid density, the specific heat, and the thermodynamic critical field should be described within similar model by using the same set of parameters. For example, this is the case in analysis of conventional elemental and binary superconductors within the framework of the empirical $\alpha-$model (see Refs.~\onlinecite{Padamsee_JLTP_1973, Johnston_SST_2013, Khasanov_Bi-II_PRB_2019, Karl_Sn_PRB_2019, Khasanov_Ga-II_PRB_2020}) and the Eliashberg approach (see the classic  reviews of Carbotte, Ref.~\onlinecite{Carbotte_RMP_1990} and Carbotte and Marsiglio, Ref.~\onlinecite{Marsiglio_book_2008}); the two-gap superconductors within the self-consistent approach by using models developed by Kogan {\it et al.}\cite{Kogan_PRB_2009, Kogan_PRB_2016} and Bussmann-Holder {\it et al.}\cite{Bussmann-Holder_EPB_2004, Bussmann-Holder_Arxiv_2009, Bussmann-Holder_CondMat_2019}

Our interest in interpreting the superfluid density, the specific heat, and the thermodynamic critical field within the single-gap and the two-gap scenarios was initiated by recent publications pointing to contradictory results for BeAu superconducting compound.\cite{Rebar_PhD-Thesis_2015, Amon_PRB_2018, Rebar_PRB_2019, Singh_PRB_2019, Khasanov_BeAu_Arxiv_2020} Indeed, Refs.~\onlinecite{Rebar_PhD-Thesis_2015, Amon_PRB_2018, Rebar_PRB_2019, Singh_PRB_2019} consider BeAu to host a single isotropic gap, which is a characteristic of predominantly $s-$wave spin-singlet pairing in the weak coupling limit. Such evidence comes from the electronic specific heat  measurements, suggesting that $C_{\rm e}(T)$ is well described within the single-gap approach with the ratio $\alpha=\Delta(0)/k_{\rm B}T_{\rm c}$ ranging from the weak-coupled BCS value $\alpha_{\rm BCS}=1.764$ up to $\alpha\simeq 1.88$.\cite{Rebar_PhD-Thesis_2015, Amon_PRB_2018, Rebar_PRB_2019, Singh_PRB_2019} Here, $\Delta(0)$, $T_{\rm c}$, and $k_{\rm B}$ are the zero-temperature value of the gap, the superconducting transition temperature, and Boltzmann constant, respectively.
On the other hand, precise measurements of the thermodynamic critical field reported in Ref.~\onlinecite{Khasanov_BeAu_Arxiv_2020} show that the temperature evolution of $B_{\rm c}$ can not be described within the single-gap scenario and the presence of at least two different types of the superconducting order parameters is required. The analysis of $B_{\rm c}(T)$ within the self-consistent two-gap approach suggests the presence of two superconducing energy gaps with the ratios $\Delta/k_{\rm B}T_{\rm c}\simeq2.26$ and $1.19$ for the big and the small gap, respectively.

The aim of the present paper is to resolve the above contradiction in interpreting the $C_{\rm e}(T)$ and $B_{\rm c}(T)$ data of BeAu. In order to solve this dilemma both, $C_{\rm e}(T)$ and $B_{\rm c}(T)$, were simultaneously analysed by means of the single-gap and the two-gap models. Our result confirms that the two-gap approach allows to describe both thermodynamic quantities with the same set of parameters. Moreover, the fact that $C_{\rm e}(T)$ could be interpreted within the single gap scenario becomes an interesting feature of BeAu superconductor. In BeAu, two contributions to the electronic specific heat, by being summed together, result in a behavior, which is indistinguishable from expectations of the single-gap scenario. The simulations show that this is also the case for the temperature evolution of the superfluid density. Our results imply, therefore, that in some particular cases, as {\it e.g.}, in the case of BeAu studied here, measurement of one single thermodynamic quantity may not be enough in order to speculate about the mechanism of superconductivity. The validity of the model used to describe the experimental data for one particular thermodynamic quantity needs to be cross checked by analysing the other quantities as well. The correct interpretation requires that various thermodynamic quantities need to be satisfactorily described within framework of the similar model with the same set of the parameters.

The paper is organized as follows: Section~\ref{sec:Theory} describes the single-gap and two-gap approaches in describing temperature dependencies of the specific heat, the supefluid density, and the thermodynamic critical field. The single-gap $\alpha-$model (Section~\ref{sec:single-gap}) and the self-consistent two-gap model (Section~\ref{sec:two-gap_model}) are considered. Section~\ref{sec:comparison_with_experiment} details the analysis of $C_{\rm e}(T)$ and $B_{\rm c}(T)$ of BeAu within the single-gap and two-gap scenarios. In Section~\ref{seq:comparison_single_vs_two-gap}, the specific heat and the superfluid density data, simulated by means of two-gap model, are compared with the one-gap model fits. Conclusions follow in Section~\ref{sec:conclusions}.

\section{The single-gap and the two-gap approaches}\label{sec:Theory}

This section describes the single-gap and the two-gap models, which are traditionally used to analyze temperature dependencies of the electronic specific heat $C_{\rm e}$, the superfluid density $\rho_{\rm s}$, and the thermodynamic critical field $B_{\rm c}$. The single-gap approach is based on the model first introduced by Padamsee {\it et al.},\cite{Padamsee_JLTP_1973} and is known as the $\alpha$-model. The model was recently reconsidered by Johnston.\cite{Johnston_SST_2013} The two-gap model for analysing $C_{\rm e}(T)$ and $\rho_{\rm c}(T)$ was introduced by Bouquet {\it et al.},\cite{Bouquet_EPL_2001} Carrington and Manzano,\cite{Carrington_2003} and Fletcher {\it et al.}\cite{Fletcher_PRL_2005} for analyzing the specific heat and the superfluid density data of MgB$_2$. The extension of this model for analyzing the thermodynamic critical field  data is given in Ref.~\onlinecite{Khasanov_BeAu_Arxiv_2020}. There are two approaches considering different temperature dependencies of the big [$\Delta_{1}(T)$] and the small [$\Delta_{2}(T)$] superconducting energy gaps. The simplest, {\it i.e.} the two-gap version of the $\alpha-$model, assumes the gaps to be described within the weak-coupled BCS approach by following the functional form derived by M\"{u}hlschlegel.\cite{Muehlschlegel_ZPhys_1959} The more advanced version determines temperature evolutions of the small and the big gap by solving the coupled gap equations self-consistently.\cite{Kogan_PRB_2009,Kogan_PRB_2016, Bussmann-Holder_EPB_2004, Bussmann-Holder_Arxiv_2009, Bussmann-Holder_CondMat_2019}

\subsection{Single-gap approach \label{sec:single-gap}}

\subsubsection{$C_{\rm e}$, $\rho_{\rm s}$, and $B_{\rm c}$ within the single-gap BCS model}

Within the isotropic $s-$wave gap scenario,  the temperature dependencies of the electronic specific heat $C_{\rm e}$, the superfluid density $\rho_{\rm s}(T)/\rho_{\rm s}(0)$, and the thermodynamic critical field $B_{\rm c}(T)$ can be obtained analytically:\cite{Tinkham_book_1975, Padamsee_JLTP_1973, Johnston_SST_2013, Khasanov_BeAu_Arxiv_2020}
\begin{eqnarray}
\frac{C_{\rm e}(T)}{\gamma_{\rm n}T_{\rm c}} &=& \frac{6}{\pi^2 k_{\rm B}^3 T_{\rm c} T^2} \int_0^{\infty}f(1-f) \nonumber \\
&& \times \left[ \epsilon^2+\Delta(T)^2- \frac{T\; \partial \Delta(T)^2}{2\;\partial T} \right] d\epsilon,
 \label{eq:Ce_single-gap}
\end{eqnarray}
\begin{equation}
\frac{\rho_{\rm s}(T)}{\rho_{\rm s}(0)}= 1 -\frac{1}{2 T}\int_{0}^{\infty} \cosh^{-2} \left[ \frac{\sqrt{\epsilon^2+\Delta(T)^2}}{2 T} \right]
{\rm d} \epsilon,
 \label{eq:Superfluid_single-gap}
\end{equation}
and
\begin{equation}
\frac{B_{\rm c}^2(T)}{\gamma_{\rm n}T_{\rm c}^2} = 8\pi \frac{F_{\rm n}-F_{\rm s}}{\gamma_{\rm n} T_{\rm c}^2}.
 \label{eq:Critical-field_single-gap}
\end{equation}
Here, $f=[1+\exp(\sqrt{\epsilon^2+\Delta(T)^2}/k_{\rm B}T)]^{-1}$ is  the Fermi function and $\gamma_{\rm n}$ is the normal state electronic specific heat coefficient (Sommerfeld constant). $F_{\rm n}$ and $F_{\rm s}$ are the normal and superconducting state free energies given by:\cite{Johnston_SST_2013, Khasanov_BeAu_Arxiv_2020}
\begin{equation}
\frac{F_{\rm n}}{\gamma_{\rm n}T_{\rm c}^2} = -\frac{ T^2}{2 T_{\rm c}^2}.
 \label{eq:Fn}
\end{equation}
and
\begin{eqnarray}
\frac{F_{\rm s}[T,\Delta(T)]}{\gamma_{\rm n}T_{\rm c}^2}  & = &-\frac{3 }{ \pi^2 k_{\rm B}^2 T_{\rm c}^2}
\bigg[  \frac{\Delta(T)^2}{4}  \bigg. \nonumber \\
&& + \left.  \int_0^\infty f\; \frac{2\varepsilon^2 + \Delta(T)^2}{E} d\varepsilon \right].
\label{eq:Fs}
\end{eqnarray}
Note that the Eqs.~\ref{eq:Ce_single-gap}, \ref{eq:Critical-field_single-gap}, \ref{eq:Fn}, and \ref{eq:Fs} are expressed in cgs units.\cite{Johnston_SST_2013}

\subsubsection{Single-gap $\alpha-$model \label{sec:alpha-model}}

\begin{figure}[htb]
\includegraphics[width=0.9\linewidth]{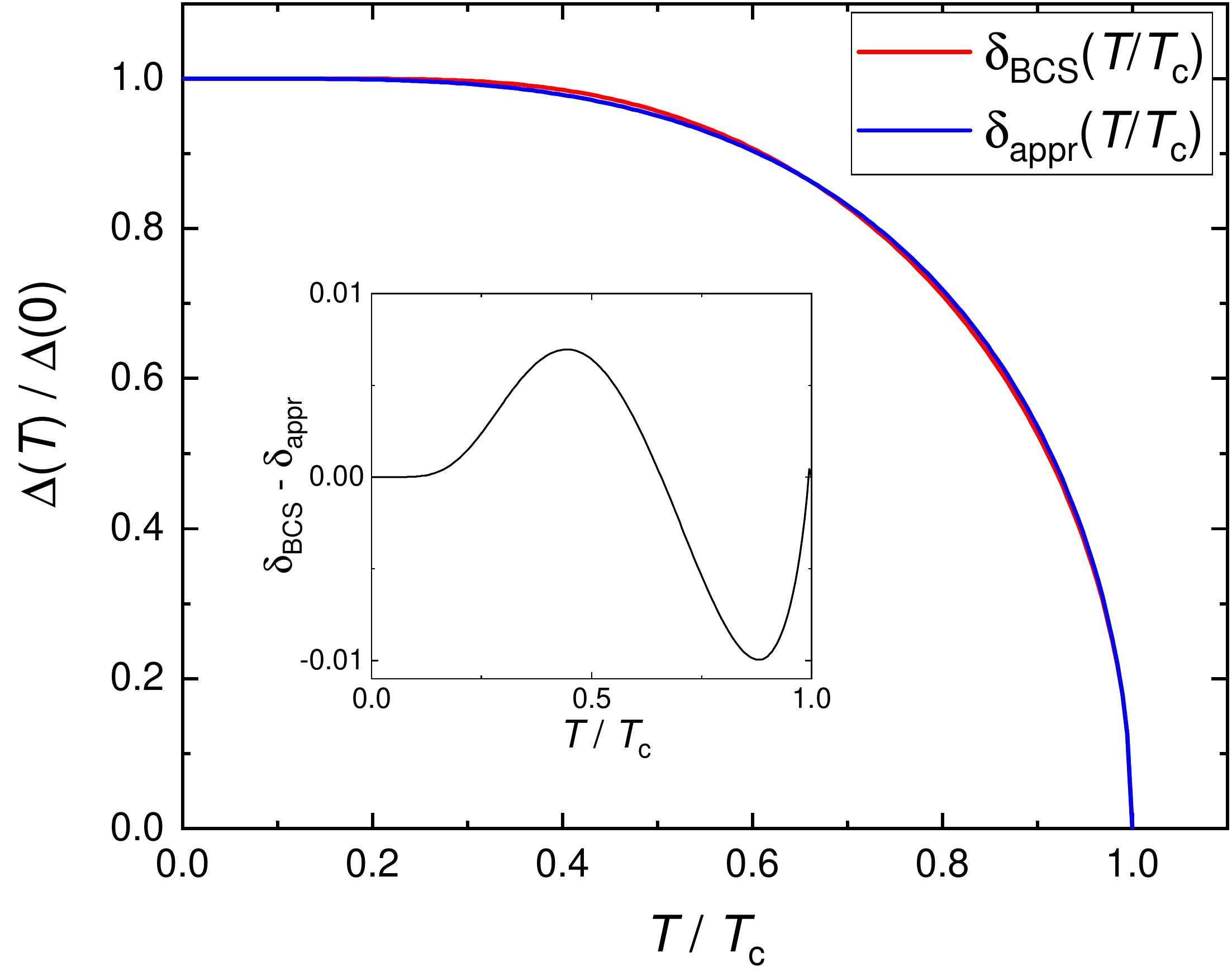}
%
\caption{ Normalized BCS energy gap $\delta_{\rm BCS}(T/T_{\rm c})=\Delta_{\rm BCS}(T/T_{\rm c})/\Delta_{\rm BCS}(0)$,  obtained by solving the self-consistent single-gap equation Eq.~\ref{eq:Self-consistent_single-gap} (red curve) and the approximated gap calculated by means of Eq.~\ref{eq:Approximated_Gap} (blue curve). The inset shows the difference $\delta_{\rm BCS}(T/T_{\rm c})-\delta_{\rm appr}(T/T_{\rm c})$ between the BCS and the approximated solutions. }
 \label{fig:BCS-gap}
\end{figure}

Originally, the $\alpha$-model was adapted from the single-band Bardeen-Cooper-Schrieffer (BCS) theory of superconductivity in order to explain deviations of the temperature behavior of the electronic specific heat and the thermodynamic critical field from the weak-coupled BCS prediction.\cite{Padamsee_JLTP_1973, Johnston_SST_2013} The model assumes that the temperature dependence of the normalized superconducting energy gap:
\begin{equation}
\delta_{\rm BCS}(T)=\frac{\Delta(T)}{\Delta(0)}=\frac{\Delta_{\rm BCS}(T)}{\Delta_{\rm BCS}(0)},
\label{eq:delta(t)}
\end{equation}
[$\Delta(0)$ is the zero-temperature value of the gap] is the same as for BCS theory.\cite{Muehlschlegel_ZPhys_1959}
The parameter
\begin{equation}
\alpha = \frac{\Delta(0)}{k_{\rm B} T_{\rm c}}
\end{equation}
is further introduced in order to account for deviation from the weak-coupled BCS prediction $\alpha_{\rm BCS}\simeq1.764$.

The BCS temperature dependence of the gap is obtained by solving the self-consistent gap equation:\cite{Tinkham_book_1975,Johnston_SST_2013}
\begin{equation}
1= NV \int_0^{\omega_{\rm D}} \frac{1}{\sqrt{\epsilon^2+\Delta(T)^2}} \tanh \frac{\epsilon^2+\Delta(T)^2}{2 k_{\rm B} T} {\rm d}\epsilon.
 \label{eq:Self-consistent_single-gap}
\end{equation}
Here $\omega_{\rm D}$ is the Debye frequency, $N$ is the normal-state electronic density of states at the Fermi level, and $V$ is the electron-phonon interacting potential. The temperature dependence of the normalized BCS gap is normally approximated by:\cite{Carrington_2003, Khasanov_LiPdB_PRB_2006, Khasanov_MoSb_PRB_2008}
\begin{equation}
\delta_{\rm appr}(T)=\tanh \{1.82[1.018(T_{\rm c}/T-1)]^{0.51} \},
 \label{eq:Approximated_Gap}
\end{equation}
which results in less than 1\% deviation from the exact solution of Eq.~\ref{eq:Self-consistent_single-gap}.

Figure~\ref{fig:BCS-gap} compares the exact solution of Eq.~\ref{eq:Self-consistent_single-gap} with the results of Eq.~\ref{eq:Approximated_Gap}. The deviation of the approximated gap function $\delta_{\rm appr}(T)$ from $\delta_{\rm BCS}(T/T_{\rm c})$ is given in the inset.

\subsection{Two-gap approach \label{sec:two-gap_model}}

\subsubsection{$C_{\rm e}$, $\rho_{\rm s}$, and $B_{\rm c}$ within the two-gap model}

Following Refs.\onlinecite{Bouquet_EPL_2001, Carrington_2003, Prozorov_SST_2006, Khasanov_La214_PRL_2007, Khasanov_Y123_PRL_2007, Khasanov_Y123_JSNM_2008, Guritanu_PRB_2004, Kogan_PRB_2009, Bussmann-Holder_Arxiv_2009, Khasanov_BeAu_Arxiv_2020}, the temperature evolutions of $C_{\rm e}$, $\rho_{\rm s}$, and $B_{\rm c}$ within the two-gap approach could be obtained as:
\begin{eqnarray}
\frac{C_{\rm e}(T)}{\gamma_{\rm n}T_{\rm c}} &= &\frac{C_{\rm e1}(T)+C_{\rm e2}(T)}{\gamma_{\rm n}T_{\rm c}} \nonumber \\
       &=& w_{C_{\rm e}} \frac{C_{\rm e1}(T)}{\gamma_{\rm n1}T_{\rm c}} + (1-w_{C_{\rm e}}) \frac{C_{\rm e2}(T)}{\gamma_{\rm n2}T_{\rm c}},
 \label{eq:Ce_two-gap}
\end{eqnarray}
\begin{eqnarray}
\frac{\rho_{\rm s}(T)}{\rho_{\rm s}(0)} &=& \frac{\rho_{\rm s1}(T)+\rho_{\rm s2}(T)}{\rho_{\rm s}(0)} \nonumber\\
    &=& w_\rho\frac{\rho_{\rm s1}(T)}{\rho_{\rm s1}(0)} + (1-w_\rho)\frac{\rho_{\rm s2}(T)}{\rho_{\rm s2}(0)},
 \label{eq:Superfluid_two-gap}
\end{eqnarray}
and
\begin{eqnarray}
\frac{B_{\rm c}^2(T)}{\gamma_{\rm n}T_{\rm c}^2} &=& 8 \pi \frac{F_{\rm n}-F_{\rm s1}-F_{\rm s2}}{\gamma_{\rm n}T_{\rm c}^2} \nonumber \\
       &=& 8 \pi \left[ \frac{F_{\rm n}}{\gamma_{\rm n}T_{\rm c}^2} - w_{F_{\rm s}}\frac{F_{\rm s1}}{\gamma_{\rm n1}T_{\rm c}^2} -(1-w_{F_{\rm s}})\frac{F_{\rm s2}}{\gamma_{\rm n2}T_{\rm c}^2} \right] . \nonumber \\
       &&
  \label{eq:Critical-field_two-gap}
\end{eqnarray}
Here 1 and 2 are gap indices. $w_{C_{\rm e}}$, $w_\rho$, and $w_{F_{\rm s}}$ are the contributions of the first gap to the resulting electronic specific heat, the superfluid density and the superconducting free energy, respectively. $\gamma_{\rm n}=\gamma_{\rm n1}+\gamma_{\rm n2}$. Note that in accordance with Eqs.~\ref{eq:Ce_two-gap} and \ref{eq:Critical-field_two-gap}
\begin{equation}
w_{C_{\rm e}} \equiv w_{F_{\rm s}} = \frac{\gamma_{n1}}{\gamma_{n1}+\gamma_{n2}}.
 \label{eq:omega_Ce_Bc}
\end{equation}

The analytical equation for $w_\rho$ was calculated by Kogan {\it et al.},\cite{Kogan_PRB_2009}, within the framework of the self-consistent two-gap model. $w_\rho$ was found to depend on the partial densities of states and the average Fermi velocities of the first and second band, respectively. In general $w_\rho\neq w_{C_{\rm e}} \equiv w_{F_{\rm s}}$.

\subsubsection{Self-consistent two-gap model}

The self-consistent two-gap model was introduced shortly after formulation of the BCS theory by Suhl {\it et al.},\cite{Suhl_PRL_1959} and Moskalenko,\cite{Moskalenko_FMM_1959} in order to account for a more complex Fermi surface topology than the one introduced by BCS weak-coupled theory. Extensions of this approach were suggested shortly after.\cite{Gelikman_SPSS_1966} The important clue of these extensions to superconductivity is an interband pair scattering potential, which leads to enhanced pair scattering via exchange through an additional channel.
The revised version of the model for two isotropic $s-$wave gaps was recently reconsidered in a series of publications by Kogan {\it et al.},\cite{Kogan_PRB_2009,Kogan_PRB_2016} and Bussmann-Holder {\it et al.}\cite{Bussmann-Holder_EPB_2004, Bussmann-Holder_Arxiv_2009, Bussmann-Holder_CondMat_2019} Here, the approach developed in Refs.~\onlinecite{Bussmann-Holder_EPB_2004, Bussmann-Holder_Arxiv_2009, Bussmann-Holder_CondMat_2019} is  described.

Following Refs.~\onlinecite{Bussmann-Holder_EPB_2004, Bussmann-Holder_Arxiv_2009, Bussmann-Holder_CondMat_2019, Khasanov_FeSe_PRL_2010, Gupta_FrontPhys_2019, Khasanov_BeAu_Arxiv_2020}, within the two-gap approach, the coupled $s$-wave gap equations are described as:
\begin{equation}
\Delta_i (T)= \sum_{j=1}^2 \int_{0}^{\omega_{D_j}}\frac{N_jV_{ij}\Delta_j(T)}{\sqrt{\epsilon^2+\Delta_j^2(T)}}\tanh \frac{\sqrt{\epsilon^2+\Delta_j^2(T)}}{2k_{\rm B}T}d\epsilon
 \label{eq:Self-consistent_model}
\end{equation}
Here $N_i$ is the partial density of states for $i-$th band at the Fermi level ($N_1+N_2=1$); $V_{ii}$ and $V_{ij\neq i}$ are the intraband and the interband interaction potentials, respectively. For simplicity, one normally assumes that the Debye frequency $\omega_{\rm D}$ is the same for both bands ($\omega_{\rm D1}=\omega_{\rm D2}=\omega_{\rm D}$).\cite{Kogan_PRB_2009, Bussmann-Holder_Arxiv_2009, Khasanov_FeSe_PRL_2010, Gupta_FrontPhys_2019, Khasanov_BeAu_Arxiv_2020}

The system of self-consistent coupled gap equations (Eq.~\ref{eq:Self-consistent_model}) might be solved for the certain values of the intraband ($V_{11}$, $V_{22}$) and interband ($V_{12}$, $V_{21}$) coupling potentials and the Debye frequency ($\omega_{\rm D}$), which results in temperature dependencies of the big $\Delta_1(T)$ and the small $\Delta_2(T)$ superconducting energy gaps. The respective $\Delta_{1}(T)$ and $\Delta_{2}(T)$ are further substituted into Eqs.~\ref{eq:Ce_two-gap}, \ref{eq:Superfluid_two-gap}, and \ref{eq:Critical-field_two-gap} (with the individual components described by Eqs.~\ref{eq:Ce_single-gap}, \ref{eq:Superfluid_single-gap}, \ref{eq:Fn}, and \ref{eq:Fs}) and the temperature dependecies of $C_{\rm e}(T)$, $\rho_{\rm s}(T)$, and $B_{\rm c}(T)$ are obtained.

\section{Comparison with the experiment \label{sec:comparison_with_experiment}}

\begin{figure*}[htb]
\includegraphics[width=0.95\linewidth]{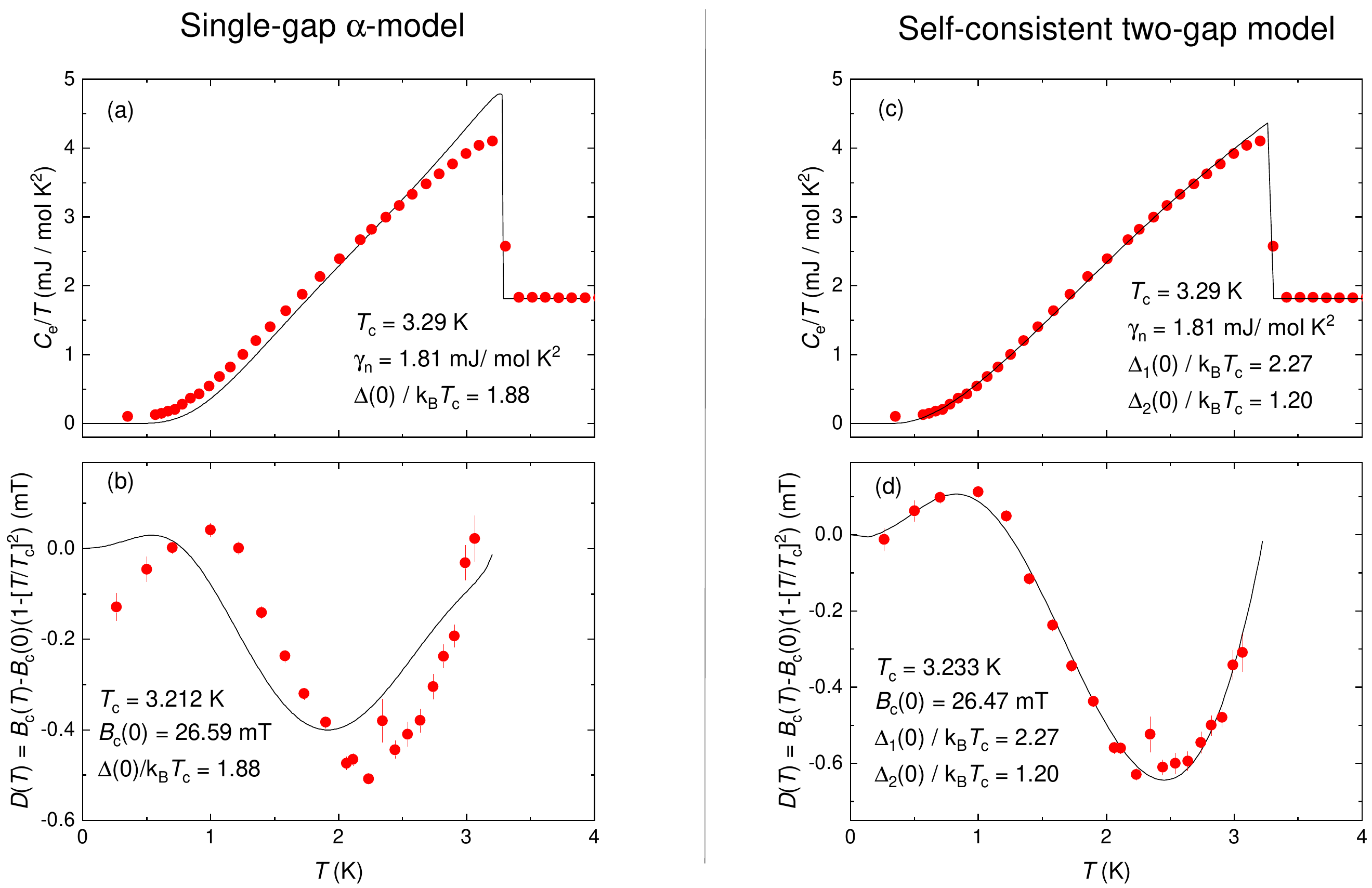}
%
\caption{(a) and (c) The electronic specific heat $C_{\rm e}(T)$ of BeAu (after Ref.~\onlinecite{Amon_PRB_2018}). (b) and (d) The deviation of the temperature evolution of the thermodynamic critical field $B_{\rm c}(T)$ of BeAu from the parabolic behaviour $D(T)=B_{\rm c}(T)-B_{\rm c}(0)[1-T^2/T_{\rm c}^2]$ (after Ref.~\onlinecite{Khasanov_BeAu_Arxiv_2020}). The solid lines in panels (a) and (b) are fits of the single-gap $\alpha-$model to the experimental data with the same $\alpha=\Delta(0)/k_{\rm B}T_{\rm c}$. The solid lines in panels (c) and (d) are fits of the self-consistent two-gap model to the experimental data with the same big $\Delta_{1}(T)$ and small $\Delta_{2}(T)$ superconducting energy gaps and the weighting factor $w_{C_{\rm e}}\equiv w_{F_{\rm s}}=0.62$. See text for details.  }
 \label{fig:models}
\end{figure*}

In this section, the comparison of the single-gap and the two-gap approaches in analyzing the temperature evolution of the specific heat $C_{\rm e}(T)$ and the thermodynamic critical field $B_{\rm c}(T)$ of BeAu superconductor is discussed.
The $C_{\rm e}(T)$ and $B_{\rm c}(T)$ data sets were taken from Refs.~\onlinecite{Amon_PRB_2018} and \onlinecite{Khasanov_BeAu_Arxiv_2020}, respectively. Note that currently three sets of $C_{\rm e}(T)$ data and one set of $B_{\rm c}(T)$ data for BeAu are available in the literature.\cite{Rebar_PhD-Thesis_2015, Amon_PRB_2018, Rebar_PRB_2019, Singh_PRB_2019, Khasanov_BeAu_Arxiv_2020} Among them, the $C_{\rm e}(T)$ and $B_{\rm c}(T)$ dependencies from Refs.~\onlinecite{Amon_PRB_2018} and \onlinecite{Khasanov_BeAu_Arxiv_2020} were measured on BeAu samples from the same grown batch. For this reason, these two data sets are going to be compared in this paper. Note that the comparison of $B_{\rm c}(T)$ from Ref.~\onlinecite{Khasanov_BeAu_Arxiv_2020} with $C_{\rm e}(T)$ curves from Refs.~\onlinecite{Rebar_PhD-Thesis_2015, Rebar_PRB_2019, Singh_PRB_2019} leads to the same basic conclusions (not shown).

Figure~\ref{fig:models} represents the results of the analysis within the single-gap $\alpha-$model and the self-consistent two-gap scenario.
Obviously, and as is already pointed out by Padamsee {\it et al.}\cite{Padamsee_JLTP_1973} in his original $\alpha-$model paper almost 50 years ago, the model, accounting correctly for the symmetry and the temperature evolution of the superconducting order parameter(s), should be able to describe various thermodynamic quantities with a similar set of parameters. Bearing this in mind, both, $C_{\rm e}(T)$ and $B_{\rm c}(T)$, data sets were analyzed simultaneously. Here-after we discuss separately the results obtained within the single- and two-gap approaches.

\subsection{$C_{\rm e}(T)$ and $B_{\rm c}(T)$ of BeAu: single-gap scenario \label{sec:analysis_single-gap}}

Within the single-gap scenario, Eqs.~\ref{eq:Ce_single-gap} and \ref{eq:Critical-field_single-gap} were fitted to the $C_{\rm e}(T)$ and $B_{\rm c}(T)$ data sets, respectively. The temperature dependence of the superconducting energy gap was assumed to be the same and was described within the $\alpha$-model (Section~\ref{sec:alpha-model}) by using
\begin{equation}
  \Delta(T)=\Delta(0)\; \delta_{\rm BCS}(T)= \alpha\; k_{\rm B} T_{\rm c}\; \delta_{\rm BCS}(T).
  \label{eq:alpha-model_analysis}
\end{equation}
The results of the analysis are shown by solid black lines in panels (a) and (b) of Fig.~\ref{fig:models}. Note that, the thermodynamic critical field data from Ref.~\onlinecite{Khasanov_BeAu_Arxiv_2020} are expressed terms of the so-called deviation function $D(T)=B_{\rm c}(T) -B_{\rm c}(0)[1-T^2/T_{\rm c}^2]$, which represents the deviation of $B_{\rm c}(T)$ from the parabolic type of behaviour. The analysis does not lead to a satisfactorily agreement between the theory and the experiment. Both theoretical $C_{\rm e}(T)$ and $D(T)$ curves deviate strongly from the data.  The best fit parameters are: $T_{\rm c}^{C_{\rm e}}=3.29$~K, $\gamma_{\rm n}=1.81$~mJ/mol K$^2$ for the specific heat, and $T_{\rm c}^{B_{\rm c}}=3.212$~K, $B_{\rm c}(0)=26.59$~mT for the thermodynamic critical field, respectively. The parameter $\alpha=\Delta(0)/k_{\rm B}T_{\rm c}$ defining the absolute value of the superconducting energy gap was found to be $\alpha\simeq 1.88$.

It is worth to note here that the single gap $\alpha-$model could be well fit to the specific heat data itself. The fit yields $\alpha=1.734$, which is only slightly smaller than the weak-coupled BCS value $\alpha_{\rm BCS}=1.764$. This agrees with the previous observations  suggesting the presence of a single isotropic energy gap in BeAu.\cite{Amon_PRB_2018, Rebar_PhD-Thesis_2015, Rebar_PRB_2019, Singh_PRB_2019} On the other hand, no agreement with $B_{\rm c}(T)$ for {\it any} $\alpha$ values was found.

\subsection{$C_{\rm e}(T)$ and $B_{\rm c}(T)$ of BeAu: two-gap scenario}

The two-gap analysis followed the procedure described in Section~\ref{sec:two-gap_model}. Eqs.~\ref{eq:Ce_two-gap} and \ref{eq:Critical-field_two-gap} were fit simultaneously to the $C_{\rm e}(T)$ and $B_{\rm c}(T)$ experimental data. The individual components of these equations were described by Eqs.~\ref{eq:Ce_single-gap}, \ref{eq:Fn}, and \ref{eq:Fs}, respectively. The temperature dependencies of the big [$\Delta_1(T)$] and the small [$\Delta_2(T)$] superconducting energy gaps were obtained by solving the system of coupled nonlinear equations (Eq.~\ref{eq:Self-consistent_model}). At a first glance, the number of the fit parameters is quite high, so the fitting procedure might be unstable and good fits are expected to be found for various combinations of the parameters. The number of the fit parameters are: 7 for Eq.~\ref{eq:Self-consistent_model} ($N_1$, $N_2$, $V_{11}$, $V_{12}$, $V_{21}$, $V_{22}$, and $\omega_{\rm D}$), 4 for Eq.~\ref{eq:Ce_two-gap} ($T_{\rm c}$, $\gamma_{\rm n}$, $\gamma_{\rm n1}$, and $\gamma_{\rm n2}$), and 5 for Eq.~\ref{eq:Critical-field_single-gap} [$T_{\rm c}$, $B_{\rm c}(0)$, $\gamma_{\rm n}$, $\gamma_{\rm n1}$, and $\gamma_{\rm n2}$]. This results in total of 16 fit parameters. In reality, the number of the fit parameters could be reduced substantially, since some of them are related to the each other, while the others could be determined in an independent set of experiments. Indeed:\\
 (i) Within the free electron approximation, the product $\gamma_{{\rm n}i}/ \gamma_{\rm n}$ ($i=1,2$) from Eqs.~\ref{eq:Ce_two-gap} and \ref{eq:Critical-field_two-gap}  is equal to the corresponding partial density of states $N_i$ in Eq.~\ref{eq:Self-consistent_model}, so that $\gamma_{\rm n1}/ \gamma_{\rm n} \equiv N_1$ and $\gamma_{\rm n2}/ \gamma_{\rm n}\equiv N_2$.\\
 (ii) The value of the Debye frequency could be independently derived from the specific heat measurements. Amon {\it et al.}\cite{Amon_PRB_2018} have found $\omega_{\rm D}\simeq 33.4$~meV. Note that slightly smaller $\omega_{\rm D}\simeq 25.4$~meV was reported by Rebar {\it et al.}\cite{Rebar_PhD-Thesis_2015,Rebar_PRB_2019} \\
 (iii) The value of the electronic specific heat component (Sommerfeld constant) could be determined from the linear fit of $C_{\rm e}(T)$ data for $T$ above $T_{\rm c}$ [see Fig.~\ref{fig:models}~(a) and (c)]. The fit reveals $\gamma_{\rm n}\simeq 1.81$ mJ/mol K$^2$.\\
 (iv) The electronic specific heat component $\gamma_{\rm n}$ is derived as a sum of two contributions $\gamma_{\rm n}=\gamma_{\rm n1}+\gamma_{\rm n2}$.\\
By considering the above arguments, the total number of the fitting parameters for simultaneous analysis of $C_{\rm e}(T)$ and $B_{\rm c}(T)$ data is reduced from 16 to  8, namely: 4 coupling constants (2 intraband: $V_{11}$ and $V_{22}$ and 2 interband: $V_{12}$ and $V_{21}$ coupling strengths), 2 transition temperatures ($T_{\rm c}^{C_{\rm e}}$, and $T_{\rm c}^{B_{\rm c}}$),\cite{comment} the zero-temperature value of the thermodynamic critical field $B_{\rm c}(0)$, and the weighting factor $w_{C_{\rm e}}\equiv w_{F_{\rm s}}=\gamma_{\rm n1}/\gamma_{\rm n}$.
It is important to emphasize here, that the above described single-gap  fit [see the discussion in Section~\ref{sec:analysis_single-gap} and Figs.~\ref{fig:models}~(a) and (b)], although it looks simpler compared to the two-gap one, requires 5 fit parameters.

\begin{figure}[htb]
\includegraphics[width=0.9\linewidth]{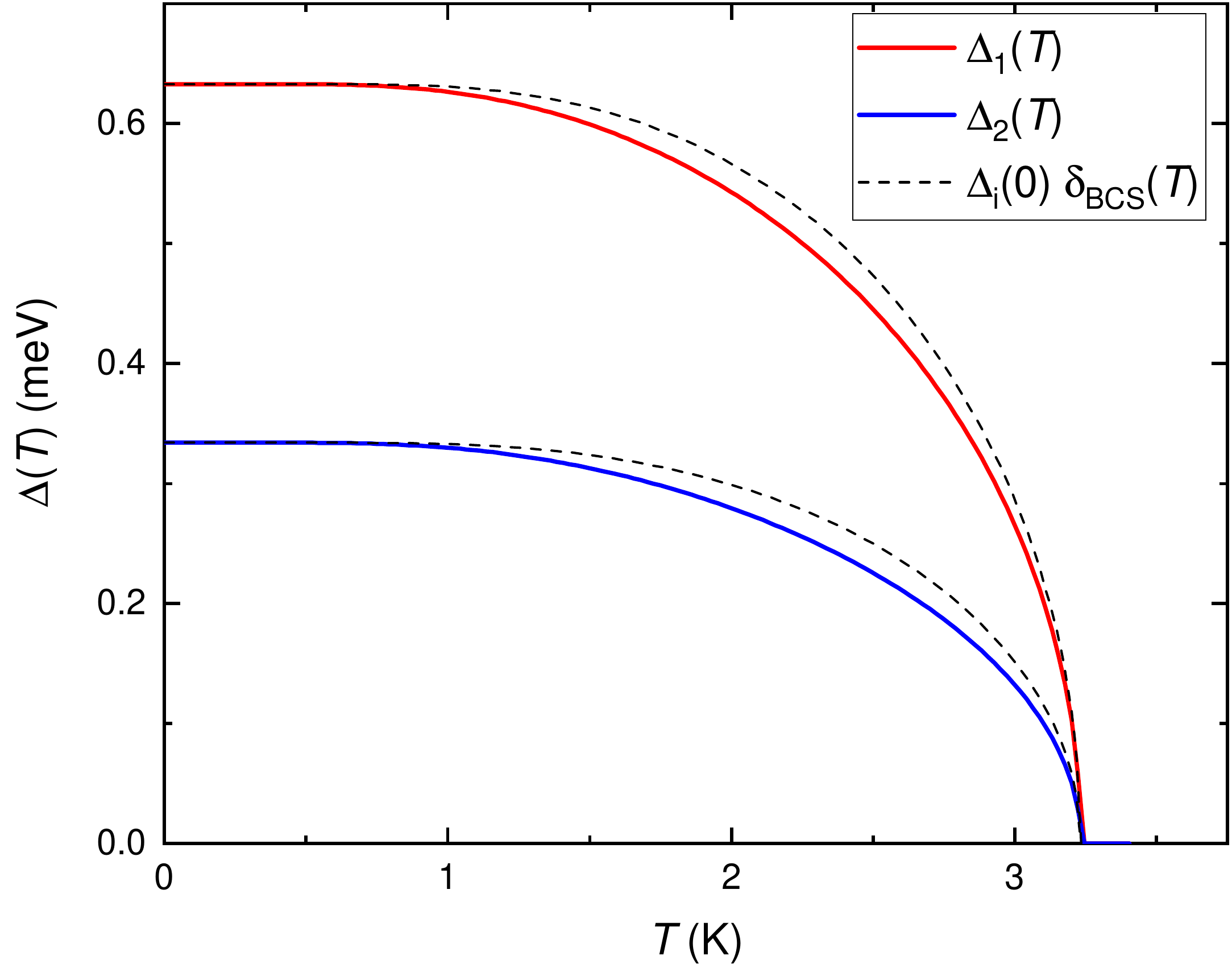}
%
\caption{ Temperature evolution of the big ($\Delta_1$, red line) and the small ($\Delta_2$, blue line) superconducting energy gaps, obtained within the framework of the self-consistent two-gap model. The dashed lines are the weak-coupled BCS prediction. }
 \label{fig:gaps}
\end{figure}

The analysis of $C_{\rm e}(T)$ and $B_{\rm c}(T)$ data within the two-gap scenario was performed in a few steps. First, by solving two nonlinear coupled gap equations (Eq.~\ref{eq:Self-consistent_model}), the temperature dependencies of the big and small superconducing gaps  were calculated. Note that the combination of the coupling strengths and the Debye frequency determines uniquely the value of $T_{\rm c}$. By substituting $\Delta_{1}(T)$ and $\Delta_{2}(T)$ into Eqs.~\ref{eq:Ce_two-gap} and \ref{eq:Critical-field_two-gap} the corresponding $C_{\rm e}(T)$ and $B_{\rm c}(T)$ were derived. The difference between the experimental data and the theory was minimized by adjusting $T_{\rm c}^{C_{\rm e}}$, $T_{\rm c}^{B_{\rm c}}$, $B_{\rm c}(0)$, and  $w_{C_{\rm e}}\equiv w_{F_{\rm s}}$. In the  next step, the coupling strengths parameters were readjusted and the calculations of $C_{\rm e}(T)$ and $B_{\rm c}(T)$ were repeated. After about 10 to 15 such iterations the fit converges. The results of the self-consistent two-gap analysis are presented in Figs.~\ref{fig:models}~(c) and (d) for $C_{\rm e}(T)$ and $B_{\rm c}(T)$ data sets, respectively.  Obviously, the self-consistent two-gap model describes both sets of the experimental data remarkably well. The fit parameters are: $B_{\rm c}(0)\simeq 26.47$~mT, $w_{C_{\rm e}}\simeq 0.622$, $V_{11}\simeq 0.243$, $V_{22}\simeq 0.312$, $V_{12}\simeq 0.281$, and $V_{21}\simeq 0.068$. The values of the superconducting transition temperature are $T_{\rm c}^{C_{\rm e}}=3.29$~K and $T_{\rm c}^{B_{\rm c}}=3.233$~K.

The temperature evolution of the big and the small gap and the comparison of their temperature dependencies with the weak-coupled BCS prediction (see Fig.~\ref{fig:BCS-gap}) are presented in Fig.~\ref{fig:gaps}. Obviously, both $\Delta_{1}(T)$ and $\Delta_{2}(T)$ are slightly weaker than the expectation of the weak-coupled BCS model (dashed lines). Such a behaviour is expected in the case of multi-band superconductors. It is confirmed theoretically as well as experimentally for the most famous two-gap superconductor MgB$_2$.\cite{Kogan_PRB_2009, Kim_Symmetry_2019, Golubov_JPCM_2002, Floris_PRL_2005, Chen_APL_2008}

\section{Comparison between the single-gap and two-gap approaches \label{seq:comparison_single_vs_two-gap}}

\begin{figure}[htb]
\includegraphics[width=0.9\linewidth]{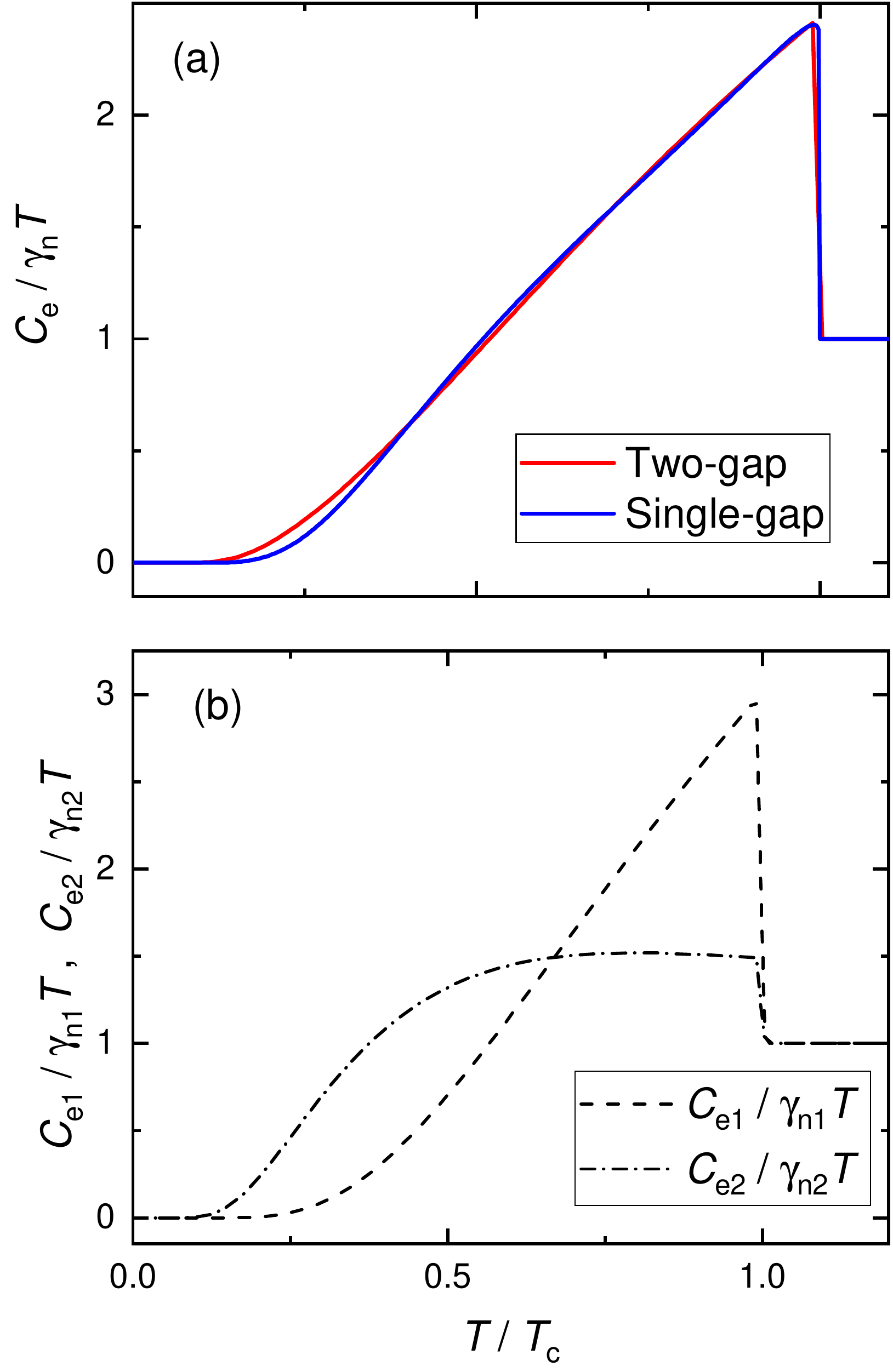}
%
\caption{(a) Comparison of the temperature evolution of the electronic specific heat obtained within the framework of the self-consistent two-gap model (red line) with the fit to the single-gap $\alpha-$model (Eq.~\ref{eq:Ce_single-gap}, blue line). (b) The individual contributions to the specific heat, obtained within the framework of the self-consistent two-gap model. See text for details.}
 \label{fig:specific-heat}
\end{figure}

\begin{figure}[htb]
\includegraphics[width=0.92\linewidth]{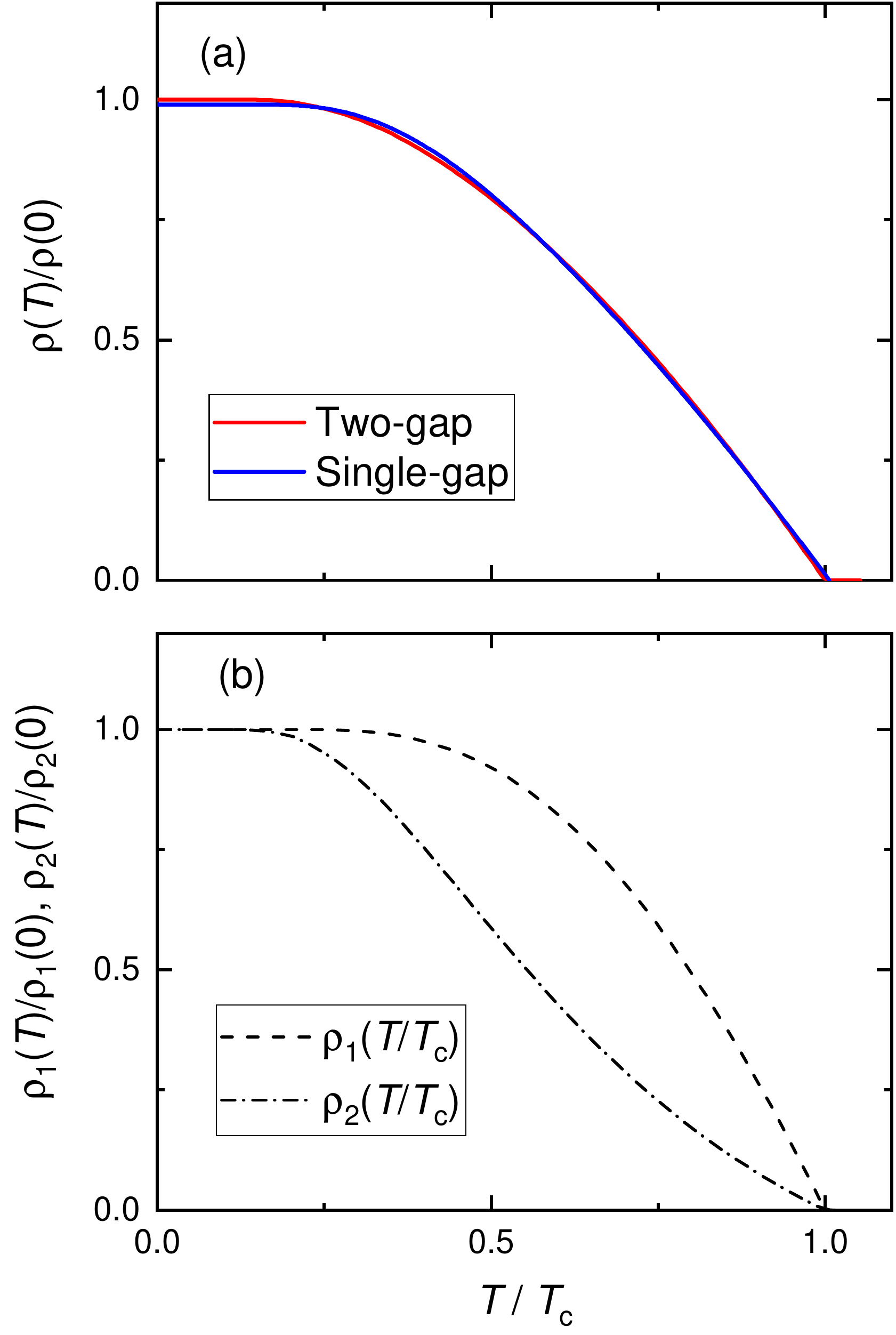}
%
\caption{ (a) Comparison of the temperature evolution of the superfluid density $\rho_{\rm s}(T)\rho_{s}(0)$ obtained within the self-consistent two-gap model (Eq.~\ref{eq:Superfluid_single-gap}, red line) with the fit to the single-gap $\alpha-$model (Eq.~\ref{eq:Superfluid_single-gap}, blue line). (b) The individual contributions to the superfluid density, obtained within the framework of the self-consistent two-gap model. See text for details.}
 \label{fig:superfluid}
\end{figure}

The obvious question arises: why the previous analysis of the specific heat data of BeAu superconductor was found to be consistent with the presence of a single isotropic energy gap.\cite{Amon_PRB_2018, Rebar_PhD-Thesis_2015, Rebar_PRB_2019, Singh_PRB_2019} In order to answer this question, the simulated 'two-gap' $C_{\rm e}(T)$ curve [see Fig.~\ref{fig:models}~(c)] was reanalyzed by using the single-gap $\alpha-$model (see Eq.~\ref{eq:Ce_single-gap} and Fig.~\ref{fig:specific-heat}). In addition, the superfluid density $\rho_{\rm s}(T)/\rho_{\rm s}(0)$ curve was calculated by means of Eq.~\ref{eq:Superfluid_two-gap} and further reanalyzed within the single-gap approach by using Eq.~\ref{eq:Superfluid_single-gap} (see Fig.~\ref{fig:superfluid}). In two-gap $\rho_{\rm s}$ calculation, the 'weighting' parameter $w_\rho$ was assumed to be equal to $w_{C_{\rm e}}$. Surprisingly, in both cases the single-gap curves stays relatively close to the two-gap ones.
The fit parameters are: $\alpha=1.734$ and 1.675 for the specific heat and the superfluid density, respectively. Note that $\alpha=1.734$, obtained from the fit of 'two-gap' specific heat data, stays relatively close to the values reported in Refs.~\onlinecite{Amon_PRB_2018, Rebar_PhD-Thesis_2015, Rebar_PRB_2019, Singh_PRB_2019}.

This is rather unexpected, meaning that in the particular case of BeAu superconductor the summed contributions of the big and the small gap [dashed and dash-dotted lines in Figs.~\ref{fig:specific-heat}~(b) and \ref{fig:superfluid}~(b)] are indistinguishable from the single-gap scenario. By following the data presented in Figs.~\ref{fig:specific-heat} and \ref{fig:superfluid}, this becomes true for at least two thermodynamical quantities, such as the electronic specific heat and the superfluid density. On the other hand, the thermodynamic critical field $B_{\rm c}(T)$ is failed to be described within the single-gap behaviour and requires the use of the self-consistent two-gap model for their description (see also Ref.~\onlinecite{Khasanov_BeAu_Arxiv_2020}).

\section{Conclusions \label{sec:conclusions}}

The temperature evolution of the electronic specific heat [$C_{\rm e}(T)$] and the thermodynamic critical field [$B_{\rm c}(T)$] of BeAu superconductor were analyzed by means of the single-gap $\alpha-$model and the self-consistent two-gap model. Our results confirm that the two-gap approach allows to describe both thermodynamic quantities with the same set of parameters. The fact that $C_{\rm e}(T)$ could be interpreted within the single-gap scenario is unique for BeAu superconductor. In this particular case, the big and the small gap contributions to the electronic specific heat result in behavior which is hardly distinguishable from the single-gap approach. A similar situation was obtined for the temperature dependence of the superfluid density $\rho_{\rm s}(T)/\rho_{s}(0)$.

Our results imply that in some particular cases, as for example, is the case for BeAu studied here, the measurement of one single thermodynamic quantity may not be enough in order to establish the symmetry of the superconducting order parameter. A successful gap analysis requires that various thermodynamic quantities be described using the same set of parameters.

\begin{acknowledgments}
This work was performed at the Swiss Muon Source (S$\mu$S), Paul Scherrer Institute (PSI, Switzerland). RK acknowledges helpful discussions with Rolf Lortz.  The work of RG is supported
by the Swiss National Science Foundation (SNF-Grant No. 200021-175935).
\end{acknowledgments}


\begin{thebibliography}{99}

\bibitem{Shoenberg_book_1952} D. Shoenberg, {\it Superconductivity}  (Cambridge University Press, Cambridge, 1952).

\bibitem{Parks_book_1969} {\it Superconductivity (in two volumes), edited by R.D. Parks}, (Dekker, New-York, 1969).

\bibitem{Tinkham_book_1975} M. Tinkham, {\it Introduction to Superconductivity} (Krieger Publishing company, Malabar, Florida, 1975).

\bibitem{Tilley-Tilley_book_1990} D.R. Tilley and J. Tilley, {\it Superfluids and Superconductivity, 3rd edition}  (IOP Publishing, Bristol and Philadelphiaa, 1990).

\bibitem{de_Gennes_book_1999} P.G. de Gennes, {\it Superconductivity in Metals and Alloys} (Perseus, Cambridge, MA, 1999).

\bibitem{Poole_book_2014} C.P. Poole, R. Prozorov, H.A. Farach, and R.J. Creswick {\it Superconductivity 3rd edition},  (Elsevier, Amsterdam, Netherlands, 2014).

\bibitem{Bouquet_EPL_2001} F. Bouquet, Y. Wang, R. A. Fisher, D. G. Hinks, J. D. Jorgensen, A. Junod and N. E. Phillips, Europhys. Lett., {\bf 56}, 856 (2001).

\bibitem{Carrington_2003} A.~Carrington and F.~Manzano, Physica~C {\bf 385}, 205 (2003).

\bibitem{Guritanu_PRB_2004} V. Guritanu, W. Goldacker, F. Bouquet, Y. Wang, R. Lortz, G. Goll, and A. Junod, Phys. Rev. B {\bf 70}, 184526 (2004).

\bibitem{Fletcher_PRL_2005} J. D. Fletcher, A. Carrington, O. J. Taylor, S. M. Kazakov, and J. Karpinski, Phys. Rev. Lett. {\bf 95}, 097005 (2005).

\bibitem{Prozorov_SST_2006} R. Prozorov and R.W. Giannetta,  Supercond. Sci. Technol. {\bf 19}, R41 (2006).

\bibitem{Khasanov_La214_PRL_2007} R. Khasanov, A. Shengelaya, A. Maisuradze, F. La Mattina, A. Bussmann-Holder, H. Keller, and K. A. M\"{u}ller, Phys. Rev. Lett. {\bf 98}, 057007 (2007).

\bibitem{Khasanov_Y123_PRL_2007} R. Khasanov, S. Str\"{a}ssle, D. Di Castro, T. Masui, S. Miyasaka, S. Tajima, A. Bussmann-Holder, and H. Keller, Phys. Rev. Lett. {\bf 99}, 237601 (2007).

\bibitem{Khasanov_Y123_JSNM_2008} R. Khasanov, A. Shengelaya, J. Karpinski, A. Bussmann-Holder, H. Keller, and K. A. M\"{u}ller, J. Supercond. Nov. Magn. {\bf 21}, 81 (2008).

\bibitem{Nakajima_PRL_2008} Y. Nakajima, T. Nakagawa, T. Tamegai, and H. Harima, Phys. Rev. Lett. {\bf 100}, 157001 (2008).

\bibitem{Evtushinsky_NJP_2009}  D. V. Evtushinsky, D. S. Inosov, V. B. Zabolotnyy, M. S. Viazovska, R. Khasanov, A. Amato, H. -H. Klauss, H. Luetkens, Ch. Niedermayer, G. L. Sun, V. Hinkov, C. T. Lin, A. Varykhalov, A. Koitzsch, M. Knupfer, B. Büchner, A. A. Kordyuk, and S. V. Borisenko, New J. Phys. {\bf 11}, 055069 (2009).

\bibitem{Khasanov_Ba122_PRL_2009} R. Khasanov, D. V. Evtushinsky, A. Amato, H. -H. Klauss, H. Luetkens, Ch. Niedermayer, B. Büchner, G. L. Sun, C. T. Lin, J. T. Park, D. S. Inosov, and V. Hinkov, Phys. Rev. Lett. {\bf 102}, 187005 (2009).

\bibitem{Khasanov_Sr122_PRL_2009} R. Khasanov, A. Maisuradze, H. Maeter, A. Kwadrin, H. Luetkens, A. Amato, W. Schnelle, H. Rosner, A. Leithe-Jasper, and H.-H. Klauss, Phys. Rev. Lett. {\bf 103}, 067010 (2009).

\bibitem{Khasanov_Bi2201_PRL_2008} R. Khasanov, T. Kondo, S. Str\"{a}ssle, D. O. G. Heron, A. Kaminski, H. Keller, S. L. Lee, and T. Takeuchi, Phys. Rev. Lett. {\bf 101}, 227002 (2008).

\bibitem{Gordon_PRB_2008} R. T. Gordon, M. D. Vannette, C. Martin, Y. Nakajima, T. Tamegai, and R. Prozorov, Phys. Rev. B {\bf 78}, 024514 (2008).


\bibitem{Khasanov_Bi2201_PRB_2009} R. Khasanov, T. Kondo, S. Str\"{a}ssle, D. O. G. Heron, A. Kaminski, H. Keller, S. L. Lee, and T. Takeuchi, Phys. Rev. B {\bf 79}, 180507(R) (2009).

\bibitem{Singh_PRB_2010} Y. Singh, C. Martin, S. L. Bud'ko, A. Ellern, R. Prozorov, and D. C. Johnston, Phys. Rev. B {\bf 82}, 144532 (2010).

\bibitem{Khasanov_Bi2201_PRB_2010} R. Khasanov, T. Kondo, M. Bendele, Y. Hamaya, A. Kaminski, S. L. Lee, S. J. Ray, and T. Takeuchi, Phys. Rev. B {\bf 82}, 020511(R) (2010).

\bibitem{Kim_PRB_2011} H. Kim, M. A. Tanatar, Yoo Jang Song, Yong Seung Kwon, and R. Prozorov, Phys. Rev. B {\bf 83}, 100502(R) (2011).

\bibitem{Chen_NJP_2013} J. Chen, L. Jiao, J. L. Zhang, Y. Chen, L. Yang, M. Nicklas, F. Steglich, and H. Q. Yuan, New J. Phys. {\bf 15}, 053005 (2013).

\bibitem{Khasanov_SrPtP_PRB_2014} R. Khasanov, A. Amato, P. K. Biswas, H. Luetkens, N. D. Zhigadlo, and B. Batlogg, Phys. Rev. B {\bf 90}, 140507(R) (2014).

\bibitem{Khasanov_FeSeInt_PRB_2016} R. Khasanov, H. Zhou, A. Amato, Z. Guguchia, E. Morenzoni, X. Dong, G. Zhang, and Z Zhao, Phys. Rev. B {\bf 93}, 224512 (2016).

\bibitem{Verchenko_PRB_2017} V.Yu. Verchenko, R. Khasanov, Z. Guguchia, A. A. Tsirlin, and A. V. Shevelkov, Phys. Rev. B {\bf 96}, 134504 (2017).

\bibitem{Khasanov_1144_PRB_2019} R. Khasanov, W.R. Meier, S.L. Bud'ko, H. Luetkens, P.C. Canfield, and A. Amato, Phys. Rev. B {\bf 99}, 140507(R) (2019).

\bibitem{Kim_Symmetry_2019} H. Kim, K. Cho, M.A. Tanatar, V. Taufour, S.K. Kim, S.L. Bud’ko, P.C. Canfield, V.G. Kogan, and R. Prozorov Symmetry {\bf 11}, 1012 (2019).

\bibitem{Khasanov_1144_PRB_2018} R. Khasanov, W. R. Meier, Y. Wu, D. Mou, S.L. Bud'ko, I. Eremin, H. Luetkens, A. Kaminski, P. C. Canfield, and A. Amato, Phys. Rev. B {\bf 97}, 140503(R) (2018).

\bibitem{Khasanov_Bi-III_PRB_2018} R. Khasanov, H. Luetkens, E. Morenzoni, G. Simutis, S. Sch\"{o}necker, A. \"{O}stlin, L. Chioncel, and A. Amato, Phys. Rev. B {\bf 98}, 140504(R) (2018).


\bibitem{Padamsee_JLTP_1973} H. Padamsee, J. E. Neighbor, and C. A. Shiffman, J. Low Temp. Phys. {\bf 12}, 387 (1973).

\bibitem{Johnston_SST_2013} D.C. Johnston,  Supercond. Sci. Technol. {\bf 26}, 115011 (2013).

\bibitem{Khasanov_Bi-II_PRB_2019} R. Khasanov, M.M. Radonji\'{c} H. Luetkens, E. Morenzoni, G. Simutis, S. Sch\"{o}necker, W.H. Appelt, A. \"{O}stlin, L. Chioncel, and A. Amato, Phys. Rev. B {\bf 99}, 174506 (2019).

\bibitem{Karl_Sn_PRB_2019} R. Karl, F. Burri, A. Amato, M. Doneg\`{a}, S. Gvasaliya, H. Luetkens, E. Morenzoni, and R. Khasanov, Phys. Rev. B {\bf 99}, 184515 (2019).

\bibitem{Khasanov_Ga-II_PRB_2020} R. Khasanov, H. Luetkens, A. Amato, and E. Morenzoni, Phys. Rev. B {\bf 101}, 054504 (2020).

\bibitem{Carbotte_RMP_1990} J. P. Carbotte, Rev. Mod. Phys. {\bf 62}, 1027 (1990).

\bibitem{Marsiglio_book_2008} F. Marsiglio and J.P. Carbotte, {\it Electron-Phonon Superconductivity}. In: Bennemann K.H., Ketterson J.B. (eds) Superconductivity. Springer, Berlin, Heidelberg (2008).


\bibitem{Kogan_PRB_2009} V. G. Kogan, C. Martin, and R. Prozorov, Phys. Rev. B {\bf 80}, 014507 (2009).

\bibitem{Kogan_PRB_2016} V. G. Kogan and R. Prozorov, Phys. Rev. B {\bf 93}, 224515 (2016).

\bibitem{Bussmann-Holder_EPB_2004} A. Bussmann-Holder, R. Micnas, and A. R. Bishop, Eur. Phys. J. B. {\bf 37}, 345 (2004).

\bibitem{Bussmann-Holder_Arxiv_2009} A. Bussmann-Holder, arXiv:0909.3603, unpublished.

\bibitem{Bussmann-Holder_CondMat_2019} A. Bussmann-Holder, H. Keller, A. Simon, and A. Bianconi, Condens. Matter {\bf 4}, 91 (2019).

\bibitem{Rebar_PhD-Thesis_2015} Drew Rebar, Exploring superconductivity in chiral structured BeAu, Ph.D. dissertation, Louisiana State University, 2015.

\bibitem{Amon_PRB_2018} A. Amon, E. Svanidze, R. Cardoso-Gil, M. N. Wilson, H. Rosner, M. Bobnar, W. Schnelle, J. W. Lynn, R. Gumeniuk, C. Hennig, G. M. Luke, H. Borrmann, A. Leithe-Jasper, and Yu. Grin, Phys. Rev. B {\bf 97}, 014501 (2018).

\bibitem{Rebar_PRB_2019} D. J. Rebar, S. M. Birnbaum, J. Singleton, M. Khan, J. C. Ball, P. W. Adams, J. Y. Chan, D. P. Young, D. A. Browne, and J. F. Di Tusa, Phys. Rev. B {\bf 99}, 094517 (2019).

\bibitem{Singh_PRB_2019} D. Singh, A. D. Hillier, and R. P. Singh, Phys. Rev. B {\bf 99}, 134509 (2019).

\bibitem{Khasanov_BeAu_Arxiv_2020} R. Khasanov, R. Gupta, D. Das, A. Amon, A. Leithe-Jasper, and E. Svanidze, arXiv:2003.11271.

\bibitem{Muehlschlegel_ZPhys_1959} B. M\"uhlschlegel, Z. Phys. {\bf 155}, 313, {\bf 1959}.

\bibitem{Khasanov_LiPdB_PRB_2006} R. Khasanov, I. L. Landau, C. Baines, F. La Mattina, A. Maisuradze, K. Togano, and H. Keller, Phys. Rev. B {\bf 73}, 214528 (2006).

\bibitem{Khasanov_MoSb_PRB_2008} R. Khasanov, P. W. Klamut, A. Shengelaya, Z. Bukowski, I. M. Savi\'{c}, C. Baines, and H. Keller, Phys. Rev. B {\bf 78}, 014502 (2008).

\bibitem{Suhl_PRL_1959} H. Suhl, B. T. Matthias, and L. R. Walker, Phys. Rev. Lett. {\bf 3}, 552 (1959).

\bibitem{Moskalenko_FMM_1959}V. A. Moskalenko, Fiz. Metal Metalloved. {\bf 8}, 503 (1959).

\bibitem{Gelikman_SPSS_1966} B. Geilikman, R. Zaitsev, and V. Z. Kresin, Sov. Phys. Solid State {\bf 9}, 524 (1966).

\bibitem{Khasanov_FeSe_PRL_2010} R. Khasanov, M. Bendele, A. Amato, K. Conder, H. Keller, H.-H. Klauss, H. Luetkens, and E. Pomjakushina, Phys. Rev. Lett. {\bf 104}, 087004 (2010).

\bibitem{Gupta_FrontPhys_2019} R. Gupta, A. Maisuradze, N.D. Zhigadlo, H. Luetkens, A. Amato, and R. Khasanov, Front. Phys. {\bf 8}, 2 (2020).


\bibitem{comment} In order to account for a possible temperature sensor calibration differences  the superconducting transition temperatures were assumed to be slighly different for the specific heat and the thermodynamic critical field experiments. Following the results of the two-gap fit, the temperature difference between two set of experiments does not exceeded $\simeq 60$~mK.

\bibitem{Golubov_JPCM_2002} A.A. Golubov, J. Kortus, O.V. Dolgov, O.Jepsen, Y. Kong, O.K. Andersen, B.J. Gibson, K. Ahn, and R.K. Kremer, J. Phys. Condens. Matter {\bf 14}, 1353 (2002).

\bibitem{Floris_PRL_2005} A. Floris, G. Profeta, N.N Lathiotakis, M L\"{u}ders, M.A.L. Marques, C. Franchini, E.K.U. Gross, A. Continenza, and S. Massidda, Phys. Rev. Lett. {\bf 94}, 037004 (2005).

\bibitem{Chen_APL_2008} K. Chen, Y. Cui, Q. Li, C.G. Zhuang, Z.K. Liu, and X.X. Xi, Appl. Phys. Lett.  {\bf 93}, 012502 (2008).

\end{thebibliography}
\end{document}